\documentclass[pre,twocolumn,english,superscriptaddress,floatfix,longbibliography]{revtex4-2}

\usepackage[normalem]{ulem}
\usepackage{graphicx}% Include figure files
\usepackage{dcolumn}% Align table columns on decimal point
\usepackage{bm}% bold math
\usepackage{xcolor}
\usepackage[dvipsnames]{xcolor}
\usepackage{amsmath}
\usepackage{amssymb}
\usepackage{lineno}
\usepackage{physics, mathtools}


\usepackage{hyperref}
\hypersetup{colorlinks=true, citecolor=blue, linkcolor=black, urlcolor=blue}

\begin{document}
%\linenumbers

 \author{Qian Cao}
% \altaffiliation[These authors contributed equally to this work.]{}
\affiliation{Department of Electrical Engineering and Computer Science, University of California Berkeley, Berkeley, CA, USA, 94720.}
\affiliation{Department of Physics, Washington University, St. Louis, Missouri 63130, USA}
\author{Zhen Bi}
\affiliation{Department of Physics, The Pennsylvania State University, University Park, Pennsylvania, 16802, USA}
\affiliation{Center for Theory of Emergent Quantum Matter$,$ Institute for Computational and Data Sciences$,$ The Pennsylvania State University$,$ University Park$,$ Pennsylvania$,$ 16802$,$ USA}
\author{Kater W. Murch}%
\email{katermurch@berkeley.edu}
\affiliation{Department of Electrical Engineering and Computer Science, University of California Berkeley, Berkeley, CA, USA, 94720.}
\affiliation{Department of Physics, University of California Berkeley, Berkeley, CA, USA, 94720.}

\date{\today}% It is always \today, today, %  but any date may be explicitly specified

%=========================================================================================%

\title{Mixed-State Symmetry-Protected Topology and Strong-to-Weak Spontaneous Symmetry Breaking in a Superconducting Qubit Array}

\begin{abstract}
We experimentally investigate how symmetry-protected topological order in a one-dimensional cluster state is transformed by measurement and decoherence in a five-qubit superconducting array. We first characterize the state's nonlocal string order and show that controlled dephasing selectively suppresses one symmetry sector while leaving the other robust, consistent with average symmetry-protected topological order. We then measure one sublattice in a tunable basis and show that the remaining qubits are driven between a long-range-entangled GHZ state and a paramagnetic state. When the measurement record is discarded, the conventional long-range correlator vanishes while a nonlinear fidelity correlator remains finite, providing a finite-size signature of strong-to-weak spontaneous symmetry breaking. These experiments demonstrate how conditioning, averaging, and decoherence reveal distinct manifestations of order encoded in the same underlying cluster state, and establish a superconducting-circuit setting for probing mixed-state symmetry and topology. 
\end{abstract}
\maketitle

\textit{Introduction}---Measurement is usually understood as something that destroys quantum information: observing a system collapses superpositions and, on average, can only reduce entanglement. Certain many-body states violate this intuition. The one-dimensional cluster state is short-range entangled, yet measuring a subset of its qubits can generate long-range entanglement \cite{Briegel2001}. This feature is the basis for measurement-based quantum computation, where measurement, rather than unitary evolution, drives the computation forward~\cite{Raussendorf2003,Else2012}. The cluster state exhibits such measurement-induced phenomena because it has a hidden order---a $\mathbb{Z}_2 \times \mathbb{Z}_2$ symmetry-protected topological (SPT) order encoded in nonlocal correlations~\cite{Else2012,Son2012,Pollmann2010,ChenSPT2011, Schuch2011}. When measurement outcomes are discarded, however, the system is left in a mixed state. This naturally raises the question of how SPT order manifests when information about part of the system or its environment is retained, averaged over, or discarded.

Recent theoretical work has introduced new forms of mixed-state order that address this question. In average symmetry-protected topological (ASPT) phases~\cite{Ma2023,Ma2025}, one considers a system coupled to an external measurement apparatus. Conditioning on the measurement record yields a quantum trajectory~\cite{Dalibard1992,Carmichael1993}. A protecting symmetry that is broken along individual quantum trajectories may remain restored after averaging over the ensemble, allowing nontrivial topological order to persist in a mixed state~\cite{Ma2023,Ma2025}. A complementary situation arises when a many-body system is partitioned into system and environment subsystems. Measuring the environment can project the remaining system into states with long-range order~\cite{Briegel2001, Jong22arxiv, Verresen2024,Verresen2023}, while tracing over the same environment can instead produce strong-to-weak spontaneous symmetry breaking (SW-SSB), in which conventional symmetry-breaking correlations vanish while nonlinear mixed-state correlations remain finite \cite{LeeJianXu2023, Lessa2025,Sala2024, Wang_SWSSB_arxiv_26}. Experiments have begun to probe pieces of this picture: cluster-state protocols have generated long-range entanglement \cite{deJong2024,Bumer2024,Kang2025,Kandala2025}, signatures of ASPT order have been observed in disordered Rydberg atom arrays \cite{Yue_2025}, and SW-SSB has recently been observed in a dephased Fermi gas \cite{https://doi.org/10.48550/arxiv.2604.16137}. However, the controlled response of SPT string order to tunable decoherence, and the connection between measurement-conditioned order and SW-SSB obtained by tracing over a cluster-state sublattice, remain largely unexplored experimentally.

In this Letter, we investigate these effects in a five-qubit superconducting array. We prepare a one-dimensional cluster state and characterize how its nonlocal string orders respond to controlled dephasing, observing that one symmetry sector is selectively suppressed while the other remains robust, consistent with the finite-size structure expected for ASPT order. We then measure one sublattice in a tunable basis and show that the remaining qubits are driven between a long-range-entangled GHZ state and a 
paramagnetic state. For the corresponding unconditioned state, obtained by discarding the measurement record, the conventional long-range correlator vanishes while the fidelity correlator remains finite, providing a finite-size signature of SW-SSB. 
Together, these experiments show how the underlying SPT correlations can manifest as string order, measurement-conditioned long-range entanglement, or mixed-state spontaneous symmetry breaking depending on which degrees of freedom are treated as the environment and how information about them is handled.

\begin{figure}[t]
\includegraphics[width=\columnwidth]{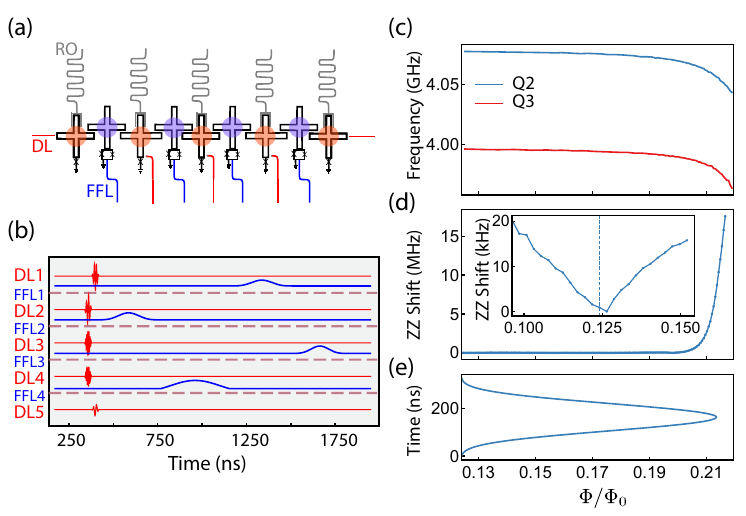}
\caption{
\textbf{Cluster state preparation in a superconducting qubit array.}
(a)~Schematic of the five-qubit device with nearest-neighbor coupling mediated by tunable couplers and dispersive readout via readout resonators. 
(b)~Pulse sequence used in preparing the one-dimensional cluster state: qubits are initialized in $|+\rangle$ states via resonant rotations (red pulses on the drive lines) and entangled via sequential nearest-neighbor controlled-$Z$ (CZ) gates (adiabatic baseband pulses on the fast flux lines). 
(c--e)~Implementation of the CZ gate using an adiabatic flux pulse.  As an example, (c) displays the dependence of $Q_2$ and $Q_3$'s frequencies on the coupler's flux. (d) Displays the measured $ZZ$ frequency shift as a function of the flux. The inset shows a zoomed-in view around the idle bias point, indicated by the vertical dashed line, where the $ZZ$ interaction between $Q_2$ and $Q_3$ is minimized. (e) Displays the adiabatic flux pulse used to implement the CZ gate between $Q_2$ and $Q_3$.  }
\label{fig:setup}
\end{figure}

\textit{Experimental platform and cluster state preparation}---We implement the experiment on a superconducting circuit comprising five fixed-frequency transmon qubits, labeled $Q_1$--$Q_5$, with transition frequencies $\omega_i/2\pi \in \{4.244,\,4.076,\,3.996,\,3.938,\,3.837\}\,\mathrm{GHz}$ [Fig.~\ref{fig:setup}(a)]. Each qubit is dispersively coupled to an individual readout resonator with frequencies $r_i/2\pi \in \{7.726,\,7.644,\,7.574,\,7.506,\,7.439\}\,\mathrm{GHz}$ and dispersive shifts $\chi_i/2\pi \in \{0.118,\,0.133,\,0.108,\,0.115,\,0.127\}\,\mathrm{MHz}$. In the dispersive regime, the interaction of a given qubit is described by $H_i/\hbar = -\chi_i a_i^\dagger a_i \sigma_z^{(i)}$, such that the resonance frequency of each cavity acquires a qubit-state-dependent shift~\cite{Blais2021}. The readout resonators are coupled to a common feedline, and the transmitted signals are amplified using a near-quantum-limited traveling-wave parametric amplifier followed by heterodyne demodulation. 

In addition to independent microwave drive lines for single-qubit control, nearest-neighbor interactions are mediated by tunable couplers implemented as auxiliary transmon elements with frequencies $c_i/2\pi \in \{4.635,4.581,4.506,4.398\}\,\mathrm{GHz}$ at their idle points~\cite{Chen2014,Yan2018,Xu2020}.  Each coupler is equipped with a dedicated fast flux line, enabling rapid tuning of its frequency and, consequently, the effective interaction between adjacent qubits. The couplers are operated near an idle point where the static interaction between qubits is minimized, suppressing residual $ZZ$ coupling and reducing unwanted phase accumulation during idle periods.

The tunable couplers modify both the neighboring qubit frequencies and their mutual interaction via off-resonant hybridization. In particular, the effective $ZZ$ interaction, defined as the conditional shift in one qubit’s transition frequency depending on the state of its neighbor, can be tuned over a wide range by varying the coupler flux. Figures~\ref{fig:setup}(c,d) show a representative characterization of the tunable coupling between $Q_2$ and $Q_3$. As the coupler flux is varied, the qubit frequencies shift due to hybridization with the coupler, while the $ZZ$ interaction is simultaneously tuned. The $ZZ$ interaction is measured by Joint Amplification of $ZZ$ interaction (JAZZ) protocol \cite{Takita2017,Li2024}. The inset indicates the idle operating point, where the residual $ZZ$ interaction is minimized to $\sim 3.2~\mathrm{kHz}$. By biasing the coupler away from this point, the interaction can be increased to $\sim 17.6~\mathrm{MHz}$, enabling fast conditional phase accumulation.

We use this tunability to implement controlled-$Z$ (CZ) gates using an adiabatic flux pulse applied to the coupler.  As shown in Fig.~\ref{fig:setup}(e), the pulse transiently increases the $ZZ$ interaction, allowing a conditional phase of $\pi$ to accumulate between the qubits while suppressing leakage and nonadiabatic transitions. The pulse is shaped using a Slepian envelope to minimize spectral broadening and residual excitations. We use interleaved randomized benchmarking \cite{Magesan2012} to the determine the four CZ gates fidelities of $0.960$, $0.923$, $0.978$ and $0.948$. The infidelity of the gate is primarily limited by leakage into the higher manifold of states of the transmon circuits.

The five-qubit cluster state is prepared by applying nearest-neighbor CZ gates to the product state $\ket{+}^{\otimes 5}$, $\ket{C_5}=\prod_{i=1}^{4}\mathrm{CZ}_{i,i+1}\ket{+}^{\otimes 5}$. It is the unique ground state of the stabilizer Hamiltonian $H_{\mathrm{cl}}=-\sum_{i=1}^{5}K_i$, with the stabilizers $K_1=X_1Z_2,\quad K_2=Z_1X_2Z_3,\quad K_3=Z_2X_3Z_4,\quad K_4=Z_3X_4Z_5,\quad K_5=Z_4X_5$.
The Hamiltonian contains three bulk stabilizers and two boundary terms. The bulk Hamiltonian, $H_{\mathrm{bulk}}=-(K_2+K_3+K_4)$, is an SPT Hamiltonian invariant under the sublattice symmetries $U_\mathrm{o}=X_1X_3X_5$ and $U_\mathrm{e}=X_2X_4$, and has a four-dimensional ground space associated with two edge qubits. The prepared state selects the edge sector by  $K_1=K_5=+1$. This boundary choice preserves $U_\mathrm{o}$, whereas the bare $U_\mathrm{e}$ maps the state to the sector $K_1=K_5=-1$. Nevertheless, the prepared state is invariant under a boundary-dressed symmetry transformation $\widetilde U_\mathrm{e}=Z_1U_\mathrm{e}Z_5=K_2K_4$; further details are provided in the End Matter. 

Rather than engineering $H_{\mathrm{cl}}$ directly, we prepare its ground state algorithmically using nearest-neighbor CZ gates on the product state $|+\rangle^{\otimes 5}$. Figure~\ref{fig:setup}(b) displays the pulse sequence used to implement this protocol, which consists of resonant $Y/2$ rotations followed by adiabatic flux pulses on the couplers. We estimate a lower-bound on the cluster-state fidelity by measuring the stabilizer generators. Defining
$P_{\mathrm{odd}} = P(K_1=K_3=K_5=+1)
$ and $
P_{\mathrm{even}} = P(K_2=K_4=+1)$,
the fidelity is lower bounded by
$F_{\mathrm{LB}} = P_{\mathrm{odd}} + P_{\mathrm{even}} - 1$~\cite{deJong2024,Tiurev2022,Tth2005}.
We obtain $F_{\mathrm{LB}} = 0.706 \pm 0.003$.  From independent calibration of the two-qubit gates, we estimate this limit to be $F_{\mathrm{CZ}}^{\mathrm{lim}} = 0.822$.

\textit{Measuring the average symmetry-protected topological string order}---The ensemble evolution of a quantum system interacting with an environment---an open quantum system---can be unraveled into pure state trajectories that are conditioned on the environment's state \cite{Dalibard1992,Carmichael1993, Murch2013}. This perspective of open quantum system dynamics motivates a distinction between \textit{exact} (strong) and \textit{average} (weak) symmetries. Consider unraveling an ensemble of quantum trajectories; an exact, or strong symmetry is one that is preserved along every individual trajectory. In contrast, an average, or weak symmetry is restored after averaging over the ensemble. This extension from individual trajectories to the ensemble level leads to the notion of an average-symmetry-protected topological (ASPT) phase. In an ASPT phase, the combination of a trajectory-level exact symmetry and an ensemble-level average symmetry can protect nontrivial topological structure in a mixed state \cite{Ma2023,Ma2025}. In the one dimensional case investigated here, this non-trivial topological structure is characterized by string order parameters which remain finite over long distances \cite{Son2011,Pollmann2010,Verresen2017}. These string order parameters can be constructed directly from the stabilizer structure of the cluster state. Multiplying stabilizers centered on the same sublattice gives the even- and odd-sublattice string orders associated with the two protecting $\mathbb{Z}_2$ symmetries;
\begin{align}
&S_\mathrm{o} = K_3=Z_2 X_3 Z_4,\\
&S_\mathrm{o}' =U_\mathrm{o}S_\mathrm{o}= K_1 K_5= X_1 Z_2 Z_4 X_5,\\
&S_\mathrm{e} = K_2 K_4= Z_1 X_2 X_4 Z_5.
\end{align}
In the absence of decoherence, the ideal cluster state has unit expectation value for all three string operators; their nonzero measured values in the prepared state provide finite-size signatures of the SPT order \cite{Briegel2001,Son2011,Else2012}.

\begin{figure}[t]
\includegraphics[width=\columnwidth]{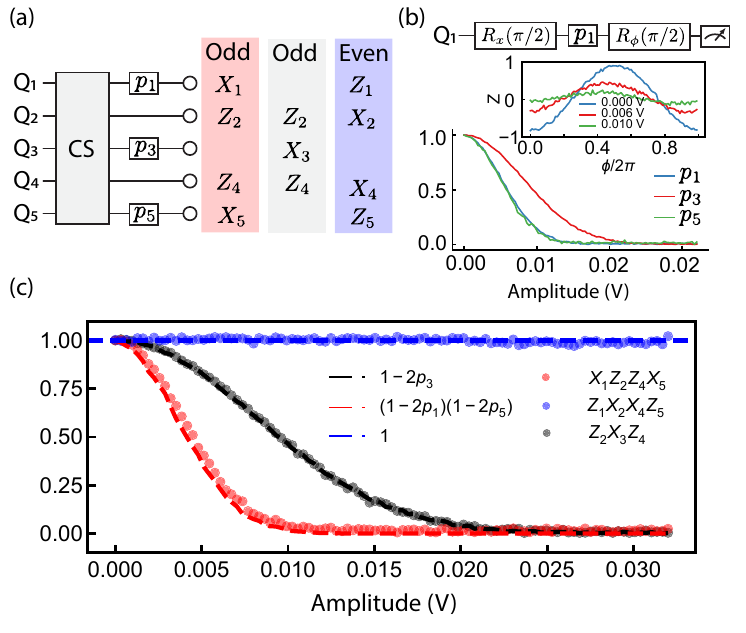}
\caption{
\textbf{Characterization of average symmetry-protected topological order under measurement-induced dephasing.}
(a) Circuit diagram for probing the response of different average string orders to dephasing strengths on $p_1$, $p_3$, $p_5$ on $Q_1$, $Q_3$, $Q_5$. (b)  Ramsey measurements are used to characterize the effective dephasing channel induced by driving the qubits' readout resonators at variable amplitude, from which a dephasing strength $p_i$ is extracted. 
(c) Measured nonlocal string order parameters as a function of measurement strength, demonstrating the persistence of SPT order over a range of conditions. The observed behavior is consistent with the selective robustness of string order expected for symmetry-protected topological phases under asymmetric decoherence.
}
\label{fig:string}
\end{figure}

To probe how the cluster-state string order responds to coupling to an external measurement apparatus, we apply weak microwave drives to the  odd sub-lattice qubits' readout resonators, as shown in Fig.~\ref{fig:string}(a). The microwave drives provide weak measurements of the odd sub-lattice qubits in the $Z$ basis \cite{Blais2021,Hatridge2013, Murch2013, Boissonneault2009}, and averaging over these measurement outcomes produces a simple dephasing channel,
\begin{equation}
\mathcal{E}_i[\rho] = (1-p_i)\rho + p_i Z_i \rho Z_i.
\end{equation}
where $p_i$ denotes the dephasing strength and corresponds to the probability of applying a Pauli-$Z$ phase flip to qubit $i$.  After applying this dephasing channel, we use projective measurements to evaluate the representative string order parameters associated with the even and odd symmetry sectors.
Figure~\ref{fig:string}(b) details the characterization of the dephasing channels $\mathcal{E}_i$.  We characterize the dephasing strength $p_i$ through a Ramsey measurement: we measure the remaining coherence as a function of the respective readout resonator's drive amplitudes. Through curve fitting, we therefore calibrate the dephasing strengths $p_i$ versus drive amplitude. 

We next measure the string correlations under increasing odd-sublattice dephasing, as shown in Fig.~\ref{fig:string}(c). We observe that the two odd-sublattice strings, $S_\mathrm{o}$ and $S_\mathrm{o}'$, decay with the dephasing strengths acting on their respective $X$ operators, whereas the even-sublattice string $S_\mathrm{e}$ remains robust. This contrasting response follows directly from the action of the dephasing channel. Stochastic $Z$ flips on the odd sublattice randomize the signs of $S_\mathrm{o}$ and $S_\mathrm{o}'$ across individual trajectories, causing their expectation values to decay after ensemble averaging. This corresponds to reducing the associated symmetry from exact to average. By contrast, the string $S_\mathrm{e}$ commutes with every Kraus operator and is therefore exactly conserved. Its robustness follows from the exact $\widetilde U_\mathrm{e}$ symmetry. This selective decay and conservation of the string orders provide a finite-size signature of the ASPT phase~\cite{Ma2023,Lee2025,Ma2025}. 

%We next measure how the average string order responds to the applied dephasing, as displayed in Fig.~\ref{fig:string}(c). We observe that the two odd strings decay approximately in proportion to the dephasing parameters acting on the corresponding qubits, while the even string remains robust to the dephasing. The decay in the odd string orders can be understand from the stochastic effect of the weak measurement along individual trajectories; the measurements induce stochastic fluctuations of the odd site's charge, breaking the trajectory-level exact (strong) $\mathbb{Z}_2$ symmetry. After averaging over trajectories, however, the resulting density matrix remains invariant under the symmetry transformation, so the symmetry is restored only as an average (weak) symmetry. This selective response of the string order demonstrates that dephasing does not uniformly destroy the underlying topological structure, but instead preserves one symmetry sector while suppressing the other, consistent with the emergence of average symmetry-protected topological order 

% The average (weak) symmetry is restored, however, after averaging over trajectories; the charge fluctuations induced by the measurement average to zero.
\textit{Measurement-induced orders and strong-to-weak spontaneous symmetry breaking}---In the measurement of the ASPT order above, we coupled the qubit array to a structured environment that corresponded to dephasing of only the odd-sublattice qubits. In this section, we change the perspective, treating the even sub-lattice as the the environment and odd sub-lattice as the quantum system of interest. We first investigate how tuning the even sub-lattice's (environment's) measurement basis drives a change in the odd-lattice order between long-range entangled and separable states. We then examine how tracing over the environment realizes the finite-size structure of SW-SSB.

\begin{figure}[t]
\includegraphics[width=.5\textwidth]{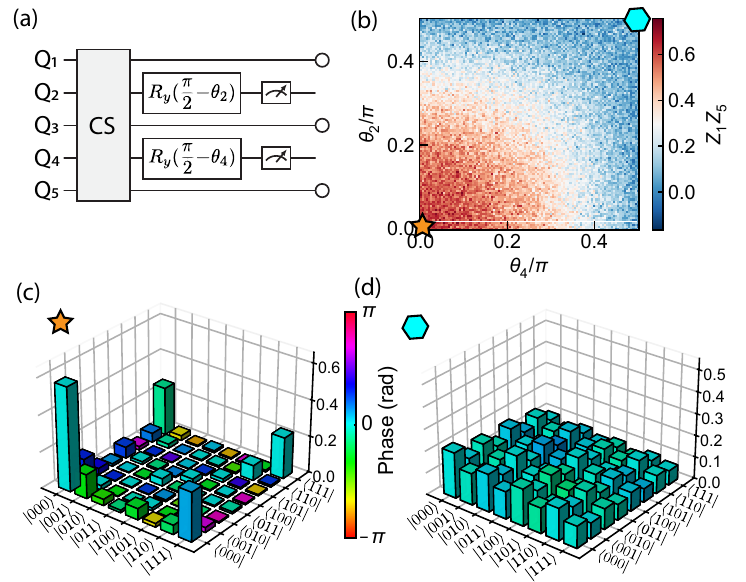}
\caption{
\textbf{Measurement-induced interpolation between long-range-entangled and separable states.} 
(a) Circuit for tuning between measurement-induced orders in the cluster state: $Q_2$ and $Q_4$ are measured in a tunable basis defined by the rotation angles $\theta_2$ and $\theta_4$, followed by state tomography conditioned on the measurement outcomes.  (b) The $Z_1 Z_5$ expectation value versus $\theta_2$ and $\theta_4$. (c) Reconstructed density matrix via conditioned quantum state tomography of $Q_1$, $Q_3$, and $Q_5$ for $\theta_2=\theta_4=0$, showing preparation of a GHZ state with 67\% fidelity. (d) Reconstructed density matrix for measurement angles  $\theta_2=\theta_4=\pi/2$ demonstrating fidelity to the paramagnetic $\ket{+}_{1,3,5}$ of 89\%.
}
\label{fig:ghz}
\end{figure}

Starting from the cluster state prepared above, we perform measurements of qubits $Q_2$ and $Q_4$ along tunable measurement axes defined as,
\begin{equation}
M_i(\theta_i)=
\cos\theta_i\,X_i+\sin\theta_i\,Z_i ,
\end{equation}
where $\theta_i$ defines the measurement basis for each qubit. The tunable measurement basis is achieved as displayed in Fig.~\ref{fig:ghz}(a). A rotation is applied to the qubits followed by projective measurement in the $Z$ basis. The measurement outcomes are denoted by $m_i$. We postselect on $(m_2,m_4)=(1,1)$ and characterize the resulting conditioned state of $Q_1$, $Q_3$, and $Q_5$. To suppress phase errors on the conditioned state, we applied an $XY$-$4$ dynamical-decoupling sequence during the measurement~\cite{Viola1999,Tripathi2022}.

By varying the measurement angle, the measurement backaction continuously changes the order of the projected state. Such measurement-induced restructuring of long-distance correlations has been studied theoretically in related many-body settings \cite{Jong22arxiv, Verresen2023, Verresen2024,LeeJianXu2023,Garratt2023,Su2024}. To quantify this behavior we measure the long-range correlator
$
\langle Z_1 Z_5\rangle ,
$
which distinguishes ferromagnetic and paramagnetic regimes. Figure~\ref{fig:ghz}(b) shows that for measurements near $\theta_2=\theta_4=0$, corresponding to the $X$ basis, the remaining qubits exhibit strong long-range correlations with $\langle Z_1 Z_5\rangle\approx 1$, indicating the few-spin limit of ferromagnetic order. In contrast, rotating the measurement basis suppresses the long-range correlations and drives the system toward a separable paramagnetic state with vanishing long-range order.

To directly visualize this evolution, we perform three-qubit quantum state tomography in representative limits. For measurements with $\theta_2,\theta_4\sim 0$, tomography of the conditioned state reveals coherence between $|000\rangle$ and $|111\rangle$, corresponding to a GHZ state and demonstrating measurement-induced long-range entanglement. This provides a finite-size realization of the general mechanism by which measurements on cluster and SPT states can generate long-range entanglement \cite{Briegel2001,Verresen2024}.  We evaluate the state fidelity between the reconstructed density matrix and the phase-corrected GHZ state
$\frac{1}{\sqrt{2}}\left(|000\rangle + e^{i\phi}|111\rangle\right)$,
where \(\phi\) captures the residual phase error. We obtain a fidelity of
$F = 0.667 \pm 0.003$,
which exceeds the 0.5 threshold required to witness genuine tripartite entanglement \cite{Toth2005}. In contrast, in the paramagnetic regime the projected state approaches
$
\ket{+}_1 \otimes \ket{+}_3\otimes \ket{+}_5,
$
which exhibits short-range correlations and negligible long-range order. We also obtain a fidelity of $F=0.888\pm0.002$.

The measurement-conditioned dynamics of the cluster state allow the remaining qubits to be tuned between long-range-entangled and separable states with distinct correlation structures. We now ask what remains of these correlations when the measurement record is discarded. Discarding the record is equivalent to tracing over $Q_2$ and $Q_4$, producing an unconditioned state $\rho_{1,3,5}$ that is independent of the measurement basis. In the $X$ basis, this state admits a particularly transparent decomposition into measurement-conditioned GHZ branches with outcome-dependent signs of the endpoint correlations. Averaging over these branches eliminates the conventional long-range order, while the resulting state retains nonlinear fidelity order characteristic of SW-SSB.

For the ideal cluster state, the reduced density matrix is
\begin{equation}
\rho_{1,3,5}=\frac{1}{8}\left(I+X_1X_3X_5\right).
\label{eq:rho135}
\end{equation}
This reduced state makes the finite-size SW-SSB structure explicit. For a density matrix $\rho$, a symmetry generated by $U$ is strong if $U\rho=e^{i\phi}\rho$, and weak if $U\rho U^\dagger=\rho$. Here, $U=X_1X_3X_5$, and the ideal reduced state satisfies the strong-symmetry condition $U\rho_{1,3,5}=\rho_{1,3,5}$ \cite{Albert2014,Bua2012}. To quantify the extent to which this symmetry is realized \emph{experimentally}, we measure the symmetry charge $\mathrm{Tr}(\rho_{1,3,5} U)\cong0.60$. Since $U$ has eigenvalues $\pm1$, this value corresponds to populations of $80\%$ and $20\%$ in the positive- and negative-charge sectors, respectively, indicating a $20\%$ leakage from the target symmetry sector. Using full state tomography, we can further construct the symmetry-resolved state by projecting the reconstructed density matrix onto the positive-charge sector \cite{Azses2020},
\begin{equation}
\rho_{+}=\frac{P_{+}\rho P_{+}}{\operatorname{Tr}(P_{+}\rho P_{+})},
\qquad P_{+}=\frac{I+U}{2}.
\end{equation}
By construction, the projected state is strongly symmetric, $U\rho_{+}=\rho_{+}$, and is supported entirely within the positive charge sector.

To discuss the symmetry breaking properties of the state, we look at the correlation functions of the charge operators $Z_i$. For the ideal state, conventional symmetry-breaking order should be absent,
\begin{equation}
\langle Z_i Z_j\rangle=\operatorname{Tr}(\rho Z_iZ_j)=0,
\end{equation}
whereas the nonlinear fidelity correlator should be one regardless of the operator seperation,
\begin{equation}
\begin{aligned}
&F^{\mathrm{SWSSB}}_{ij}\!\left(\rho,Z_iZ_j\rho Z_iZ_j\right)=1,
\end{aligned}
\end{equation}
where $F(\rho,\sigma)=\operatorname{Tr}\sqrt{\sqrt{\rho}\,\sigma\sqrt{\rho}}$. The coexistence of a vanishing ordinary correlator and a finite fidelity correlator defines the SW-SSB order~\cite{LeeJianXu2023,Lessa2025,Sala2024}. Table~\ref{tab:swssb-correlators} summarizes the measured conventional and fidelity correlators for both the unprojected and projected states. The suppression of the conventional correlators, together with the near-unity fidelity correlators in both states, supports a finite-size signature of SW-SSB.

\begin{table}[htbp]
\centering
\renewcommand{\arraystretch}{1.3}
\setlength{\tabcolsep}{1pt}
\begin{tabular}{c|ccc|ccc}
\hline\hline
State
& $\langle Z_1Z_5\rangle$
& $\langle Z_1Z_3\rangle$
& $\langle Z_3Z_5\rangle$
& $F^{\mathrm{SWSSB}}_{15}$
& $F^{\mathrm{SWSSB}}_{13}$
& $F^{\mathrm{SWSSB}}_{35}$ \\
\hline
Unprojected
& 0.053 & 0.018 & 0.025
& 0.949 & 0.978 & 0.956 \\
Projected
& 0.033 & 0.010 & $-0.007$
& 0.985 & 0.994 & 0.988 \\
\hline\hline
\end{tabular}
\caption{Experimental measurement of the conventional and fidelity correlators for the unprojected and symmetry-projected states.}
\label{tab:swssb-correlators}
\end{table}

\textit{Outlook}---
In this letter we have investigated how order encoded in a cluster state can manifest in different forms through controlled measurement and decoherence. First, odd-sublattice dephasing reduced the corresponding symmetry from exact to average, causing the associated odd string correlations to decay, while the complementary even string order remained exactly conserved. This selective response provides a finite-size probe of the underlying ASPT order. Second, we showed how directly measuring one sublattice and conditioning on the measurement outcomes produced a controlled evolution between an GHZ-ordered and a paramagnetic state. Finally we investigated how the measurement outcomes yield a mixed state with a vanishing linear correlator but a finite fidelity correlator, providing a finite-size signature of SW-SSB. These observations give a unified interpretation in terms of both the system--environment partition and the information retained about the environment. In the first part, dephasing can be viewed as entangling the system with environmental degrees of freedom and subsequently discarding, or equivalently averaging over, the environmental record. In the second part, the measured qubits form the effective environment for the remaining subsystem. Retaining their measurement outcomes resolves the remaining system into measurement-conditioned branches, whereas discarding the record produces an ensemble-averaged mixed state with a vanishing conventional correlator but a finite nonlinear fidelity correlator, analogous to Edwards--Anderson order in spin glasses \cite{Lessa2025,Sala2024,Wang_SWSSB_arxiv_26}.  From this perspective, measurement and decoherence do not simply weaken the underlying cluster-state order; rather, both the choice of which degrees of freedom are treated as the environment and whether their information is retained determine how that order is manifested, as ASPT order, measurement-conditioned GHZ order, or SW-SSB order.

These results suggest several directions for extending the connection between measurement, symmetry, and mixed-state order. First, larger qubit arrays would enable finite-size scaling of string and fidelity correlators, distinguishing finite-size signatures from genuine mixed-state phases. A central experimental challenge is to develop scalable probes of nonlinear fidelity correlations without full state reconstruction \cite{Weinstein2025,Zhang2025,7p5x-7yqb}. Second, real-time monitoring and feedback could resolve individual quantum trajectories and track the evolution of symmetry charges \cite{PhysRevX.12.041002, Barratt2022}, opening a route to measurement- and feedback-induced phase transitions \cite{Li2019,Li2018, Skinner2019,Sierant2023, ODea2024,Ravindranath2023, Friedman2023, Wu2026,Cemin25,Iadecola2023}. Repeated adaptive measurements could further be used to engineer nonequilibrium many-body states, including quantum-critical states \cite{PRXQuantum.3.040337,Lu2023}.

\textit{Acknowledgments}--- 
This work received support from the National Science Foundation award No.~PHY-2408932 and ONR Grant No.~N000142512160. Portions of this work were performed at the Aspen Center for Physics, which is supported by the National Science Foundation grant PHY-2210452. The qubit device was fabricated and provided by the Superconducting Qubits at Lincoln Laboratory (SQUILL) Foundry at MIT Lincoln Laboratory, with funding from the Laboratory for Physical Sciences (LPS) Qubit Collaboratory. ZB acknowledges the support from NSF through Grant DMR-2339319. ZB also acknowledges partial support from a Quantum SuperSEED grant (ICDS\_QS25\_029093) from the Institute for Computational and Data Sciences at the Pennsylvania State University. 

\textit{Data availability}---—The data and code that support the findings of this article are openly available at \url{https://doi.org/10.5281/zenodo.22033124}.
% \bibliographystyle{unsrt_withcaps}
% \bibliographystyle{apsrev4-2}
% \bibliography{refs_TAQS,bibfile}
\bibliography{refs_clean}%, refs_pt} 

\section*{End matter}

\subsection{Boundary choice and string order}

The experiment realizes an open-chain cluster state corresponding to a particular edge sector in the four-dimensional ground space of the $\mathbb Z_2\times\mathbb Z_2$-protected SPT Hamiltonian on an open chain \cite{Son2012}.  To make this boundary choice explicit, consider an odd-length chain $N=2L+1$ with stabilizers
\begin{equation}
 K_1=X_1Z_2,\quad K_j=Z_{j-1}X_jZ_{j+1},\quad K_N=Z_{N-1}X_N .
 \label{eq:em_stabilizers}
\end{equation}
Here the middle expression applies for $2\leq j\leq N-1$.  The bulk Hamiltonian $H_{\rm bulk}=-\sum_{j=2}^{N-1}K_j$ has the onsite symmetries
\begin{equation}
 U_\mathrm{o}=\prod_{m=0}^{L}X_{2m+1},\qquad U_\mathrm{e}=\prod_{m=1}^{L}X_{2m},
 \label{eq:em_symmetries}
\end{equation}
and a four-dimensional ground space associated with two edge qubits.  The CZ circuit prepares
\begin{equation}
 \ket{C_N}=\prod_{j=1}^{N-1}{\rm CZ}_{j,j+1}\ket{+}^{\otimes N},
 \qquad K_j\ket{C_N}=\ket{C_N},
 \label{eq:em_state}
\end{equation}
thereby selecting the boundary sector $K_1=K_N=+1$.  The endpoint stabilizers locally and independently fix the two edge degrees of freedom.  This choice preserves $U_\mathrm{o}$, whereas $U_\mathrm{e}$ anticommutes with both endpoint stabilizers and maps the prepared state into the sector $K_1=K_N=-1$.  The absence of $U_\mathrm{e}$ invariance is therefore a property of the selected boundary sector, rather than a breaking of the bulk symmetry: $H_{\rm bulk}$ still preserves the full $\mathbb Z_2^\mathrm{o}\times\mathbb Z_2^\mathrm{e}$ symmetry and its SPT structure.

Although the selected state is not invariant under the bare $U_\mathrm{e}$, completing its action with endpoint operators gives a stabilizer of the state,
\begin{equation}
 \widetilde U_\mathrm{e}=Z_1U_\mathrm{e}Z_N
 =Z_1\!\left(\prod_{m=1}^{L}X_{2m}\right)\!Z_N
 =\prod_{m=1}^{L}K_{2m}.
 \label{eq:em_dressed_symmetry}
\end{equation}
Thus $\widetilde U_\mathrm{e}$ may be viewed as the endpoint-dressed even-sublattice symmetry operator associated with the open-chain geometry.  

The role of the boundary choice is also transparent in the wavefunction.  Writing the odd spins in the $Z$-basis and the even spins in the $X$-basis gives
\begin{equation}
 \begin{aligned}
 \ket{C_N}={}&2^{-(L+1)/2}\!\sum_{\{z_{2m+1}\}}
 \ket{z_1,z_3,\ldots,z_N}_\mathrm{o}^{Z}\\
 &\otimes\bigotimes_{m=1}^{L}
 \ket{x_{2m}=z_{2m-1}z_{2m+1}}_{2m}^{X}.
 \end{aligned}
 \label{eq:em_decorated_wavefunction}
\end{equation}
Here $z_j,x_j=\pm1$ denote Pauli eigenvalues.  For this edge sector, the boundary variables $z_1$ and $z_N$ remain part of the equal-weight superposition and are not pinned in the $Z$-basis; other choices of edge state may instead fix or correlate them.  For every odd-spin configuration, each intervening even spin records whether its neighbors agree or form a domain wall. This is the manifestation of the decorated domain wall structure of the SPT wavefunction~\cite{ChenSPT2014}. 

Multiplying the local constraint $x_{2m}=z_{2m-1}z_{2m+1}$ over an interval cancels all intermediate odd-spin variables and leaves only the endpoints.  This gives the even-sublattice string
\begin{equation}
 S_\mathrm{e}[a,b]=Z_{2a-1}\!\left(\prod_{m=a}^{b}X_{2m}\right)\!Z_{2b+1}
 =\prod_{m=a}^{b}K_{2m},
 \label{eq:em_even_string}
\end{equation}
with $S_\mathrm{e}[1,L]=\widetilde U_\mathrm{e}$.  The complementary odd-sublattice string is
\begin{equation}
 S_\mathrm{o}[a,b]=Z_{2a}\!\left(\prod_{m=a}^{b}X_{2m+1}\right)\!Z_{2b+2}
 =\prod_{m=a}^{b}K_{2m+1}.
 \label{eq:em_odd_string}
\end{equation}
For the prepared state, both strings have unit expectation value for every allowed interval.  These nonlocal correlations encode the decorated-domain-wall structure characteristic of the cluster SPT.

The experiment implements dephasing on the odd sublattice,
\begin{equation}
 \mathcal E=\mathcal E_N\circ\mathcal E_{N-2}\circ\cdots\circ\mathcal E_1,
 \qquad \mathcal E_j(\rho)=(1-p_j)\rho+p_jZ_j\rho Z_j.
 \label{eq:em_channel}
\end{equation}
In the Heisenberg picture, every Kraus operator is a product of odd-sublattice $Z$ operators.  Since $S_\mathrm{e}[a,b]$ contains $Z$ on odd sites and $X$ on even sites, it commutes with every Kraus operator, and hence
\begin{equation}
 \mathcal E^\dagger(S_\mathrm{e}[a,b])=S_\mathrm{e}[a,b].
 \label{eq:em_even_conservation}
\end{equation}
Thus $S_\mathrm{e}[a,b]$ (in particular $\widetilde U_\mathrm{e}$) is an exactly conserved observable of the channel: its expectation value is unchanged in the output mixed state for any input $\rho$.  

By contrast, $S_\mathrm{o}[a,b]$ contains $X$ operators on the dephased sublattice and obeys
\begin{equation}
 \mathcal E^\dagger(S_\mathrm{o}[a,b])=
 \left[\prod_{m=a}^{b}(1-2p_{2m+1})\right]S_\mathrm{o}[a,b].
 \label{eq:em_odd_decay}
\end{equation}
Each dephased odd site contributes a factor $1-2p_j$.  For uniform $0<p<1$, $S_\mathrm{o}$ therefore decays exponentially with interval length.  

\subsection{Device setup} 
\label{app:device}
The five-qubit array is characterized by the parameters summarized in Table~\ref{tab:qubit_parameters}.  For qubit $Q_i$, $\omega_{\mathrm{q},i}/2\pi$ and $\eta_i/2\pi$ denote the qubit transition frequency and anharmonicity, respectively, while $T_{1,i}$, $T_{2,i}^*$, and $T_{2,i}^\mathrm{echo}$ characterize energy relaxation, Ramsey dephasing, and Hahn-echo coherence. The corresponding readout resonator is characterized by its resonance frequency $\omega_{\mathrm{r},i}/2\pi$, linewidth $\kappa_{\mathrm{r},i}/2\pi$, and dispersive shift $\chi_i/2\pi$. For the dispersive readout described in the main text, a microwave probe tone near $\omega_{\mathrm{r}}$ populates the resonator with an average photon number $\bar{n}$, giving an approximate measurement rate $\Gamma_\mathrm{m}\approx\frac{8\chi^2\bar{n}}{\kappa_\mathrm{r}}$.

\begin{table}[htbp]
    \centering
    \caption{Qubit parameters.}
    \label{tab:qubit_parameters}
    \renewcommand{\arraystretch}{1.3}
    \setlength{\tabcolsep}{7pt}
    \begin{tabular}{c|ccccc}
        \hline\hline
         & $Q_1$ & $Q_2$ & $Q_3$ & $Q_4$ & $Q_5$ \\
        \hline
        $\omega_{\mathrm{q},i}/2\pi$ (GHz)
        &4.244 &4.076 &3.996 &3.938 &3.837 \\

        $\eta_i/2\pi$ (MHz)
        &$-145$ &$-144$ &$-146$ &$-147$ &$-148$ \\

        $T_{1,i}$ ($\mu$s)
        &25 &59 &60 &62 &19 \\

        $T_{2,i}^*$ ($\mu$s)
        &24 &27 &30 &27 &30 \\

        $T_{2,i}^\mathrm{echo}$ ($\mu$s)
        &38 &105 &85 &92 & 38 \\

        $\omega_{\mathrm{r},i}/2\pi$ (GHz)
        &7.726 &7.644 &7.574 &7.506 &7.439 \\

        $\kappa_{\mathrm{r},i}/2\pi$ (MHz)
        &0.566 &1.067 &2.113 &1.511 &1.387 \\

        $\chi_i/2\pi$ (MHz)
        &0.118 &0.133 &0.108 &0.115 &0.127 \\
        \hline\hline
    \end{tabular}
\end{table}

\subsection{Readout error mitigation}
We perform high-fidelity single-shot readout on the selected qubits. For each qubit $q$, we independently calibrate its single-qubit assignment matrix $\beta_q$, where $(\beta_q)_{m,s}=P(m|s)$ denotes the probability of assigning the readout outcome $m\in\{0,1\}$ when the prepared state is $s\in\{0,1\}$. The calibrated single-qubit assignment matrices are

\begin{equation}
\begin{aligned}
\beta_{q_1} &=
\begin{pmatrix}
0.9738 & 0.0262 \\
0.0667 & 0.9333
\end{pmatrix},
&
\beta_{q_2} &=
\begin{pmatrix}
0.9931 & 0.0069 \\
0.0198 & 0.9802
\end{pmatrix},\\[1em]
\beta_{q_3} &=
\begin{pmatrix}
0.9878 & 0.0122 \\
0.0429 & 0.9571
\end{pmatrix},
&
\beta_{q_4} &=
\begin{pmatrix}
0.9937 & 0.0063 \\
0.0243 & 0.9757
\end{pmatrix},\\[1em]
\beta_{q_5} &=
\begin{pmatrix}
0.9870 & 0.0130 \\
0.0649 & 0.9351
\end{pmatrix}.
\end{aligned}
\end{equation}

Assuming that readout errors are independent across qubits, the assignment matrix for an arbitrary set of measured qubits $\mathcal{Q}$ is constructed as $\beta_{\mathcal{Q}}=\bigotimes_{q\in\mathcal{Q}}\beta_q ,$
where the tensor-product order is chosen consistently with the ordering of the measured probability vector. The measured probability distribution is then related to the ideal distribution by
$p_{\mathrm{meas}}=\beta_{\mathcal{Q}}p_{\mathrm{true}} .$
To mitigate readout errors, we apply an iterative Bayesian unfolding (IBU) method using the calibrated assignment matrix $\beta_{\mathcal{Q}}$ \cite{Nachman2020}.

\subsection{Measurement-induced phase correction}

In the string-order measurements, weak microwave drives are applied to the readout resonators of the odd-sublattice qubits \(Q_1\), \(Q_3\), and \(Q_5\). In addition to inducing dephasing, these drives produce measurement-induced ac-Stark shifts, which lead to additional phase accumulation of the corresponding qubits. In the laboratory frame, such a phase shift is equivalent to a rotation about the \(Z\)-axis and therefore mixes the \(X\)- and \(Y\)-components of the measured observables.

To make the extracted string order insensitive to this phase rotation, we measure the rotationally invariant magnitude of the correlators in the \(XY\) plane. For the string operator involving $Q_3$, we calculate
\begin{equation}
\left\langle S_\mathrm{o}'\right\rangle
=
\sqrt{
\left\langle Z_2 X_3 Z_4 \right\rangle^2
+
\left\langle Z_2 Y_3 Z_4 \right\rangle^2
}.
\end{equation}

Similarly, for the string operator involving  $Q_1$ and $Q_5$, we calculate
\begin{align}
\left\langle S_\mathrm{o}\right\rangle
=
\Big(
&\left\langle X_1 Z_2 Z_4 X_5 \right\rangle^2
+
\left\langle X_1 Z_2 Z_4 Y_5 \right\rangle^2
\nonumber\\
&+
\left\langle Y_1 Z_2 Z_4 X_5 \right\rangle^2
+
\left\langle Y_1 Z_2 Z_4 Y_5 \right\rangle^2
\Big)^{1/2}.
\end{align}
These combinations are invariant under arbitrary local \(Z\) rotations of the driven odd-sublattice qubits and therefore remove the effect of measurement-induced phase accumulation from the extracted string-order magnitude.

\end{document}